\documentclass[aps,prb,showpacs,twocolumn,superscriptaddress]{revtex4}
\usepackage[english]{babel}
\usepackage{graphicx}
\usepackage{color}
\usepackage[utf8]{inputenc}
\usepackage{pstricks,pst-grad,color}
\usepackage{graphicx,pifont,amssymb}
\usepackage{amsmath,amssymb}

\begin{document}

\title{Charge order in Kondo lattice systems}

\author{Robert Peters}
\email[]{peters@scphys.kyoto-u.ac.jp}
\affiliation{Department of Physics, Kyoto University, Kyoto 606-8502, Japan}

\author{Shintaro Hoshino}
\affiliation{Department of Physics, Tohoku University, Sendai
  980-8578, Japan}
\author{Norio Kawakami}
\affiliation{Department of Physics, Kyoto University, Kyoto 606-8502,
  Japan}
\author{Junya Otsuki}
\affiliation{Department of Physics, Tohoku University, Sendai 980-8578, Japan}
\affiliation{Theoretical Physics III, Center for Electronic
  Correlations and Magnetism, Institute of Physics, University of Augsburg, D-86135 Augsburg, Germany}

\author{Yoshio Kuramoto}
\affiliation{Department of Physics, Tohoku University, Sendai 980-8578, Japan}

\date{\today}


\begin{abstract}
Charge order is a commonly observed phenomenon in strongly correlated
materials. However, most theories are based on a repulsive inter-site
Coulomb interaction in order to explain charge order. We here show
that only due to local interactions, charge order is favorable in
heavy fermion systems at quarter filling. The driving force is the
Kondo effect, which leads to a 
non-linear dependence  on the filling of the lattice
site for the energy gain. We use dynamical mean field theory combined
with numerical 
renormalization group to demonstrate the existence of charge order at
quarter filling for a Kondo lattice model. 
\end{abstract}


\maketitle

\section{Introduction}
Heavy fermions are one of the most studied materials in condensed
matter physics. This is due to a large number of intriguing phenomena
ranging from quantum criticality to unconventional superconductivity,
which can be found in these
systems.\cite{coleman2007,coleman2005,gegenwart2008} Many of these
phenomena are caused by a competition between the Kondo effect and long range
order, as visualized in the Doniach phase diagram.\cite{doniach77} Most
of the long-range 
ordered phases, which have been studied theoretically so far in these
materials, are magnetically 
ordered phases like antiferromagnetism or ferromagnetism. However,
besides these phases charge order has also been found experimentally
in heavy fermion systems, e.g. in R$_5$Ir$_4$Si$_{10}$ (R = Lu or
Yb).\cite{hossain2005} These compounds exhibit a charge density wave
(CDW) at high temperatures, and a cooperation between the CDW and a
magnetic or superconducting state at much lower temperatures.
Charge order can be naturally expected in systems at quarter
filling for strong enough repulsive inter-site Coulomb interaction. This
repulsive inter-site Coulomb interaction will lead to a situation, where it is
less favorable for electrons to occupy neighboring sites equally than
forming a CDW. The formation of a CDW thus minimizes the effect of the
repulsive inter-site interaction.  
A similar effect can be observed by introducing a repulsive
{\it c}- {\it f}-electron interaction in heavy fermion models.

We will show here that it is possible to stabilize a CDW
in the Kondo lattice model without any inter-site Coulomb
interaction. As for many other interesting effects in heavy fermion
systems, the driving force is the Kondo effect. Due to the 
exponential dependence of the energy, which is gained by the Kondo effect, it
can be energetically favorable for the system to form a CDW. The
possibility of a CDW in the Kondo 
lattice model was already proposed by \textcite{Hirsch1984} in a 
strong-coupling expansion, who showed that perturbation theory creates
an effective repulsive inter-site interaction, which may lead to a CDW phase.
Recently, \textcite{otsuki2009} have shown that the static
susceptibility for a CDW diverges at quarter filling for small 
coupling strengths in the Kondo lattice model, indicating the
existence of a CDW.
Furthermore, phases with cooperation of spin and charge
order have been reported for
heavy fermion models on geometrically frustrated
lattices, and are known as partial Kondo
screened phases.\cite{Motome2010,Hayami2011,Hayami2012,Ishizuka2012,Akagi2012}
In these cases, some sites in the unit cell form a magnetically
ordered state, while others stay magnetically disordered, but may
have a different occupation number.

The purpose of this paper is to show that a pure CDW can
  be obtained at quarter filling without frustration under the 
  assumption of a bipartite lattice, and determine
  the physical properties of this phase.
Furthermore, we will 
  propose an explanation for the stability of a CDW in terms of the
  Kondo effect. 

For studying the CDW in heavy
fermions, we perform calculations for the Kondo lattice model.
The Kondo lattice model describes a lattice of local moments
which are coupled to itinerant electrons. The Hamiltonian can
be written as\cite{doniach77,lacroix1979,fazekas1991} 
\begin{displaymath}
H=t\sum_{<i,j>\sigma}c^\dagger_{i\sigma}c_{j\sigma}+J\sum_i\vec{S}_i\cdot\vec{s}_i,
\end{displaymath}
where the first term describes electron hopping on a lattice and the
second term the spin-spin interaction between a localized moment
$\vec{S}$ and the spin of a conduction electron,
$\vec{s}_i=c_{i\rho}^\dagger\vec{\sigma}_{\rho\rho^\prime}c_{i\rho^\prime}$
(with Pauli matrices $\vec{\sigma}$). Throughout the paper we will
assume an antiferromagnetic coupling, $J>0$, between the conduction electrons
and the localized spins, which is appropriate to describe heavy
fermion systems.

For solving the Kondo lattice model we use the single
  site dynamical mean field
theory (DMFT).\cite{georges1996,pruschke1995} DMFT relates the lattice
model to a quantum impurity model, 
which must be solved self-consistently. As a consequence of this
mapping onto an impurity model, the momentum dependence of the self-energy
in the obtained solution is neglected.
This approximation becomes exact in infinite
dimensions. 
However, comparison to experiments has shown that for
three dimensional systems DMFT represents a good approximation, which
is able to explain at least qualitatively many effects seen in real
experiments. To solve the effective quantum impurity model, we use the
numerical renormalization group (NRG).\cite{Wilson1975,Bulla2008} NRG
is able to accurately
calculate spectral functions for arbitrary parameters and
wide range of temperatures,\cite{peters2006,weichselbaum2007} and is also
able to resolve 
exponentially small energy scales, which might emerge in quantum
impurity models.

This paper is organized as follows: In the next
section, we will show the existence of the CDW state at quarter
filling by analyzing solutions without magnetic order.
 This will allow us to directly study the
properties of the charge ordered state. In the third section, we
allow magnetic and charge order at quarter filling simultaneously, and
analyze the cooperation/competition between these two forms of long
range order. 

\section{paramagnetic charge-density-wave at quarter filling}
The phase diagram of the Kondo lattice model within DMFT has been
investigated by a number 
of authors before.\cite{jarrell1993,sun1993,rozenberg1995,sun2005,peters2007,leo2008,otsuki2009,hoshino2010,peters2012}
  Around quarter filling, most previous studies showed
the existence of a magnetically long range ordered state.
However, recent quantum Monte Carlo calculations of static
susceptibilities by \textcite{otsuki2009} have shown the divergence of the
CDW susceptibility around quarter filling.

In order to go beyond the study of static susceptibilities and to
investigate the properties of the charge ordered state at quarter filling, we
perform calculations for an infinite dimensional Bethe lattice with
bandwidth $W$ and DOS $\rho(\omega)=8/(\pi W^2) \sqrt{W^2/4-\omega^2}$. Furthermore, we
divide the lattice into two sublattices, from now on called A- and
B-sublattice. In order to calculate the CDW state, we initialize these two
sublattices with different chemical potentials in the first DMFT
iteration, and then
iteratively solve the DMFT self-consistency equation, using the same
chemical potential for both lattice sites.
Furthermore, breaking the spin symmetry by applying
a magnetic field in the first DMFT iteration will allow to study
combinations of charge  
order and magnetism. Preserving the spin symmetry will lead to a
paramagnetic state. 

\begin{figure}[tb]
\includegraphics[width=\linewidth]{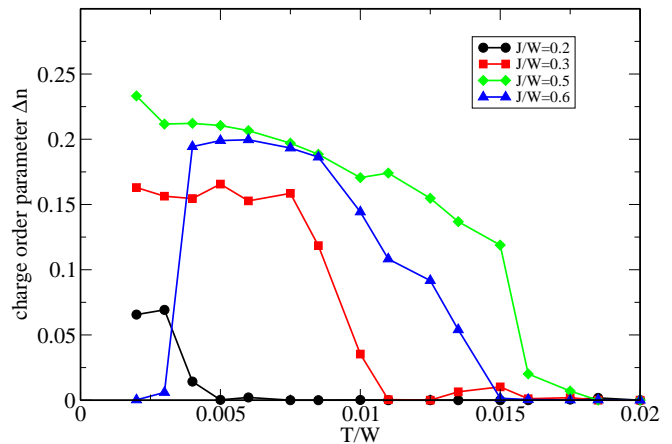}
\caption{(Color online) Charge order parameter for a paramagnetic
  state at quarter filling for different temperatures and  
  coupling strengths $J$. Lines are a guide for the eye.
 \label{fig1}}
\end{figure}
\begin{figure}[tb]
\includegraphics[width=\linewidth]{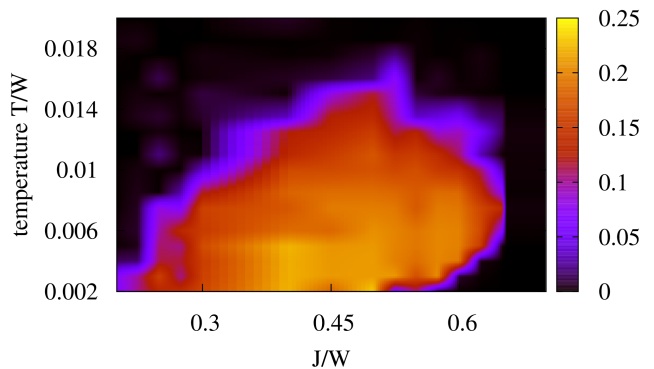}
\caption{(Color online) Magnitude of the charge order parameter for
  different temperatures and coupling strengths. 
 \label{fig2}}
\end{figure}

In this section, we neglect magnetically ordered states.
We show in Fig. \ref{fig1} that a CDW exists for the
Kondo lattice model even without inter-site repulsive Coulomb
interaction. All solutions exhibiting charge order, which are
found in this study, are located in the vicinity of
quarter filling, $\langle n_A\rangle+\langle n_B\rangle=1$.
In Fig. \ref{fig1} we show the charge order parameter $\Delta
n=\vert\langle n_A\rangle-\langle n_B\rangle\vert$ for different
temperatures and coupling strengths. Furthermore, a phase diagram
plotting the amplitude of the CDW over temperature
and coupling strength is shown in Fig. \ref{fig2}. 
Increasing the coupling strength $J$, the transition temperature as well
as the charge order parameter itself increases. The magnitude of the
charge order  reaches a maximum of $\vert n_A-n_B\vert\approx 0.23$
for $J\approx 0.5W$. Thus, the less occupied lattice sites are even at
$T=0$ filled by $\langle n\rangle\approx 0.38$.
At weak coupling strengths, $J/W<0.3$, the transition temperature increases
rapidly with increasing coupling strength. Such a rapid change in the
transition temperature could be explained by the 
exponential dependence of the Kondo effect on the coupling strength. Further
evidence that the Kondo effect is essential for 
stabilizing charge order at quarter filling is that we find no sign of
charge order for the ferromagnetically coupled Kondo lattice model, $J<0$,
for which the Kondo effect is not present.

At a coupling strength $J\approx (1/2)W$, the
  behavior changes dramatically. The transition temperature and the
  amplitude of the order parameter begin  to decrease. More importantly, 
  charge order seems to disappear at $T=0$. For coupling strengths
  $0.5W<J<0.65W$ we find a reentrant behavior  (homogeneous - charge
  order - homogeneous) for decreasing the temperature. Eventually, for
  $J\approx 0.65W$ charge order vanishes completely for all temperatures. 

\begin{figure}[tb]
\includegraphics[clip,width=\linewidth]{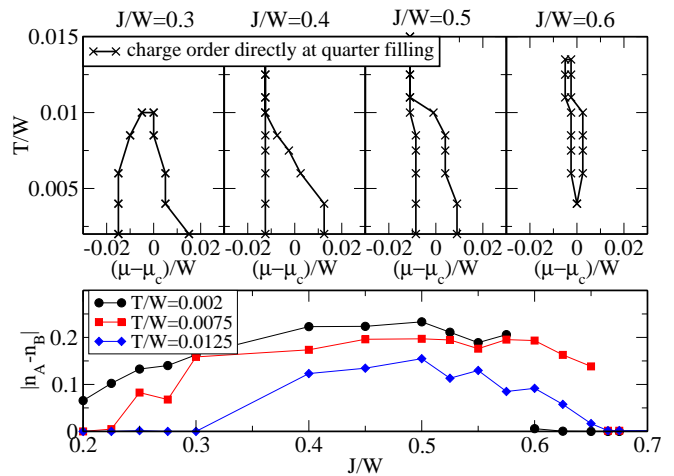}
\caption{(Color online) Upper panels: Transition temperature over
  chemical potential for the CDW exactly at quarter filling.  This plot
  illustrates the shrinking area of chemical potentials with
  increasing coupling strength, which stabilize the CDW at quarter
  filling. $\mu_c$ is the chemical potential at the center of the CDW
  phase at quarter filling.
Lower panel: Charge order
  parameter for different interaction strengths.
 \label{fig3}}
\end{figure}

To clarify the behavior at strong coupling
strengths in more detail, we show in Fig. \ref{fig3} the order
parameter at fixed temperature in the lower panel and 
the region of chemical potential,
for which a charge
ordered state can be found at quarter filling, in the upper panel.
It is evident from Fig.~\ref{fig3} that the width of the parameter region,
in which the CDW can be found, decreases with increasing coupling
strength. 
The amplitude of the charge order within this region is only slightly
changed when increasing the coupling strength. However, at a
critical coupling strength the parameter region vanishes completely.
This behavior indicates a first
  order phase transition at $T=0$.  
The corresponding jump in the
  order parameter is shown in the lower panel of Fig. \ref{fig3}.
 At high temperatures, on the
  other hand, the order parameter decreases gradually to zero,
  indicating a second order phase transition. Thus, at the critical
  interaction strength and finite temperatures, the phase transition
  changes from first to second order.

For $J/W>0.55$, we do not find a CDW at $T=0$. Only at
temperatures $T/W>0.004$, we can stabilize charge 
order. As we will show in Fig. \ref{fig6}, the charge ordered state
possesses a gap at the Fermi energy.
The region of chemical potential, in which the charge ordered
state can be stabilized at quarter filling, is supposed to be
related to the gap width, at least at $T=0$. This is because the state
does not change when changing the chemical potential within this
region.
Therefore, the shrinking of the parameter region corresponds to a
decrease of the gap width
near the critical coupling strength. 
Furthermore, we see that the gap is closed for $J/W=0.6$ for $T/W<0.004$.
Even if there would be
a tiny region of chemical potential stabilizing charge order for
$J/W\approx 0.6$ at $T=0$, 
the gap width of the state would be tiny and this state would be
unstable towards perturbations. Thus, we believe that a reentrant
behavior occurs for strong coupling strength in the vicinity of a
first order phase transition.

Comparing to experiments,
  reasonable interaction strengths are $J/W<0.3$. If we assume a
  bandwidth of $W=1eV$ in our calculations, the resulting Kondo temperature as
  well as the transition temperature of the charge order for these
  interaction strengths are $T<100K$, which are realistic temperatures
  for {\it f}-electron systems.

\begin{figure}[tb]
\includegraphics[width=\linewidth]{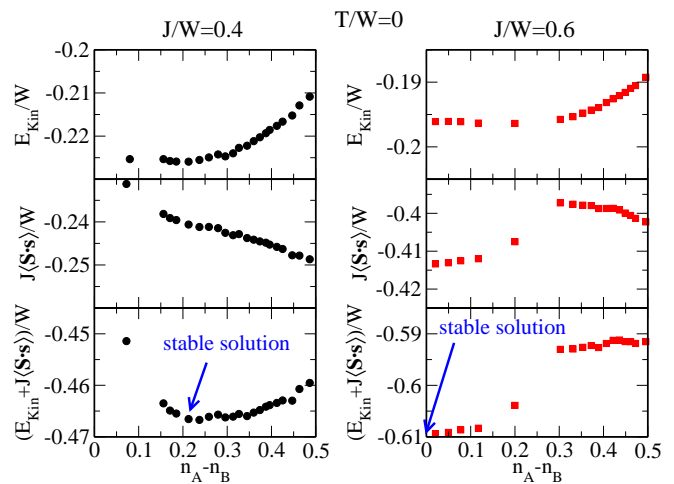}
\caption{(Color online) Kinetic energy and $J\langle \vec S \cdot \vec
  s\rangle$ (coupling energy between localized spin and conduction
  electrons) for two different interaction values $J/W$. By
  introducing a staggered potential for the sublattices A and B, we
  artificially create meta-stable solution with different occupation
  numbers, $n_A$ and $n_B$. For all solutions the system is kept
  at quarter filling. 
 \label{fig4}}
\end{figure}

J.E. Hirsch\cite{Hirsch1984} already pointed out that there exists a
nearest-neighbor repulsive density-density interaction in second-order
strong coupling expansion, 
which for weak coupling might be sufficient to drive the system
into a CDW. However, the applicability of a strong coupling theory to
explain the existence of the CDW in our calculations for small
coupling strengths is questionable.

Another way of understanding the formation
of a CDW at quarter filling is to consider the energy gain by the
Kondo effect.
In Fig. \ref{fig4}, we show the kinetic energy, $E_{kin}$, and the
energy of the exchange interaction, $J\langle \vec S \cdot \vec
  s\rangle$, for two coupling strengths calculated within
  DMFT at $T=0$. Starting from the converged DMFT solutions for both
  interaction values, we artificially change the magnitude of the
  charge order by introducing a staggered chemical potential. This
  staggered chemical potential increases or decreases the difference
  in the occupation numbers of sublattice A and B. For each obtained
  solution, we calculate the kinetic energy and the exchange
  energy. The calculated kinetic energies for both 
  interaction values exhibit minima for $n_A=n_B$, corresponding to
solutions without charge order. This can be expected, because
neglecting the spin-electron coupling, the kinetic energy is
supposed to drive the system into a
paramagnetic state. On the other hand, the exchange energy shows a
very interesting behavior. For $J/W=0.4$, which 
gives a CDW solution in Fig. \ref{fig1}, $J\langle
\vec S \cdot \vec s\rangle$ decreases for increasing difference of the
occupation numbers, $n_A-n_B$. Therefore, the exchange energy favors for 
$J/W=0.4$ a charge ordered state. The sum of kinetic energy and
exchange energy results in a minimal energy for
$n_A-n_B\approx 0.2$, which corresponds to the stable solution in our DMFT
calculations without staggered potential. In contrast to $J/W=0.4$,
the exchange energy
for $J/W=0.6$ looks completely different. This energy has two
local minima, one for a  charge ordered state and the other for a
state without charge
order. The global minimum, and thus the energetically favored state,
corresponds to the state without 
charge order. The sum of kinetic energy and exchange energy exhibits
its global minimum for solutions without charge 
order, $n_A-n_B=0$.

Thus, we arrive at the conclusion that a change in the
  behavior of the  
exchange energy, $J\langle
\vec S \cdot \vec s\rangle$, as a function of the particle number 
is responsible for driving the system into a charge ordered state.
We can qualitatively explain this change by analyzing the Kondo energy,
$E_{Kondo}\sim -W\exp(-1/(J\rho_{\epsilon_F}))$.\cite{hewson1997} 
The Kondo energy depends via the DOS on the particle number.
Thus, there might be parameter regions, where it is
energetically favorable to have different electron
occupations on sublattice A and B in 
order to gain Kondo energy. 
The importance of the Kondo effect and its dependence on the particle
number is confirmed by the following two observations:
DMFT calculations performed
for the ferromagnetically coupled Kondo lattice model, $J<0$, at
similar coupling strengths, do not show any sign
of a CDW. Furthermore, DMFT calculations performed for the
antiferromagnetically coupled Kondo lattice, $J>0$, but
having a constant DOS, do also not show
any sign of charge order. If the lattice has a constant DOS, the Kondo
energy does not depend on the particle number, so that there is no energy
gain by having different particle 
numbers on the two sublattices.

\begin{figure}[tb]
\includegraphics[width=\linewidth]{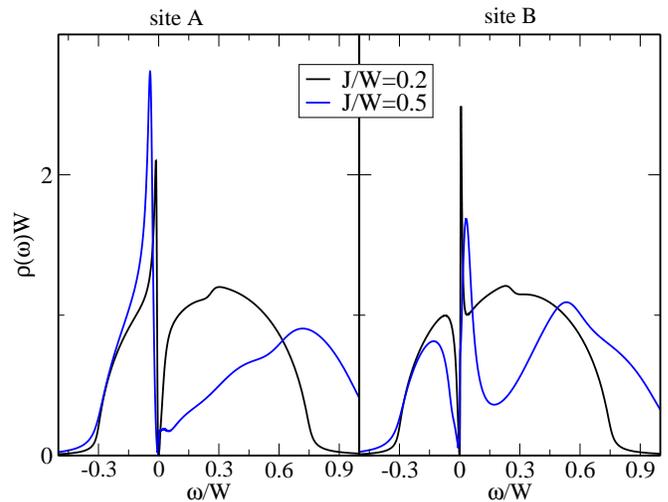}
\caption{(Color online) Spectral function of the charge ordered state
  for A- and B-sublattice for two different coupling strengths. The
  Fermi energy corresponds to $\omega=0$.
 \label{fig5}}
\end{figure}

\begin{figure*}[tb]
\includegraphics[width=0.3\linewidth]{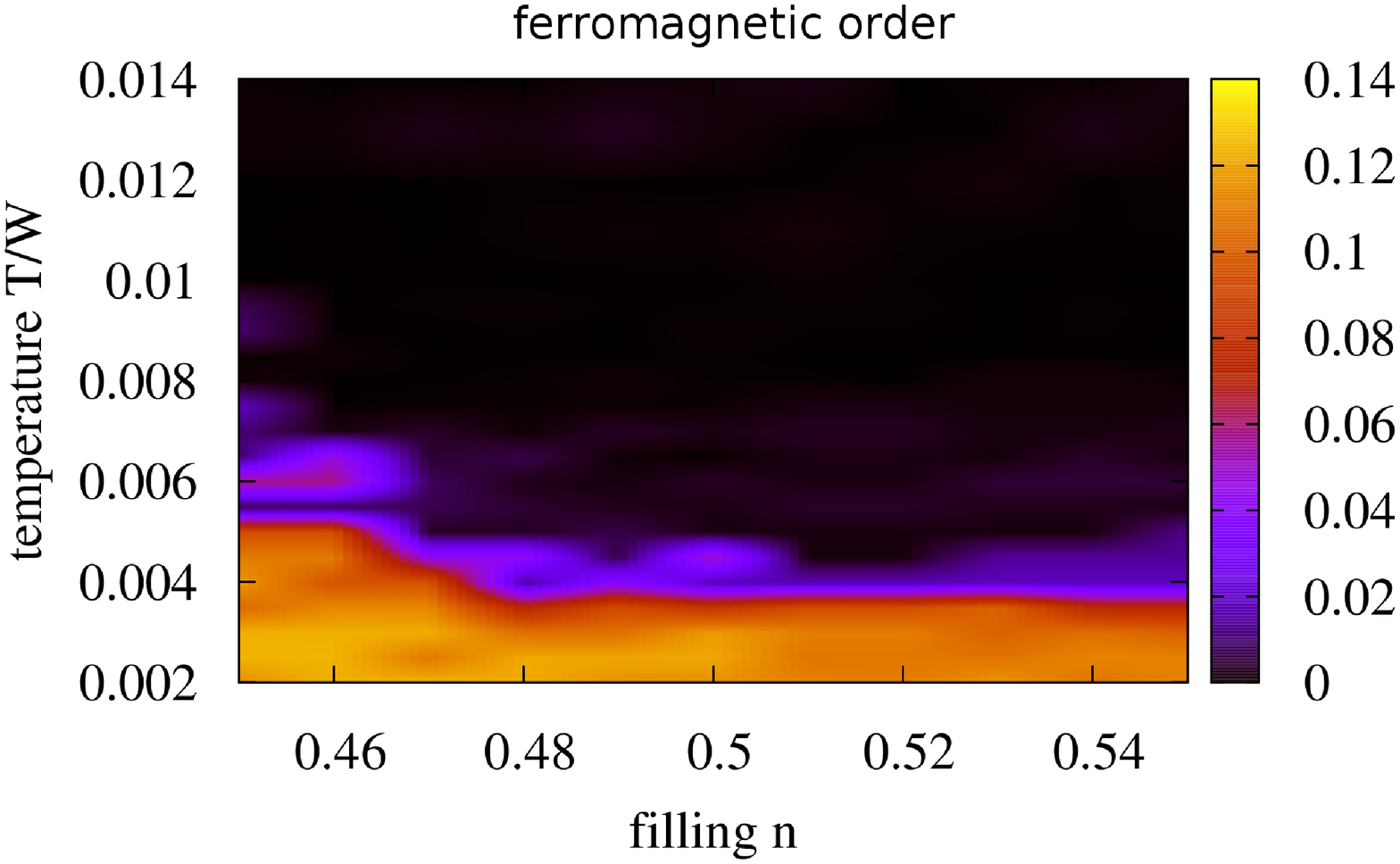}
\includegraphics[width=0.3\linewidth]{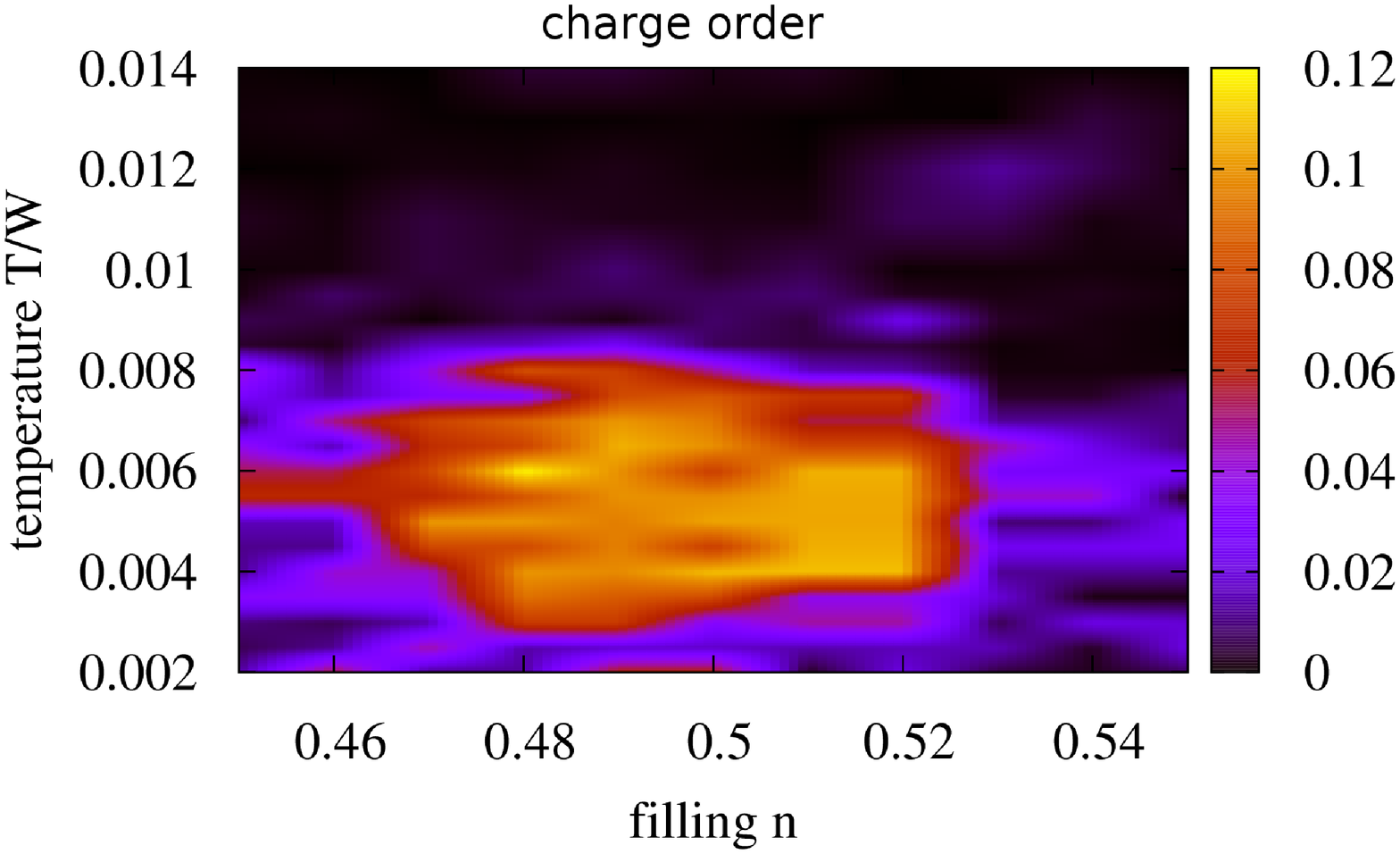}
\includegraphics[width=0.3\linewidth]{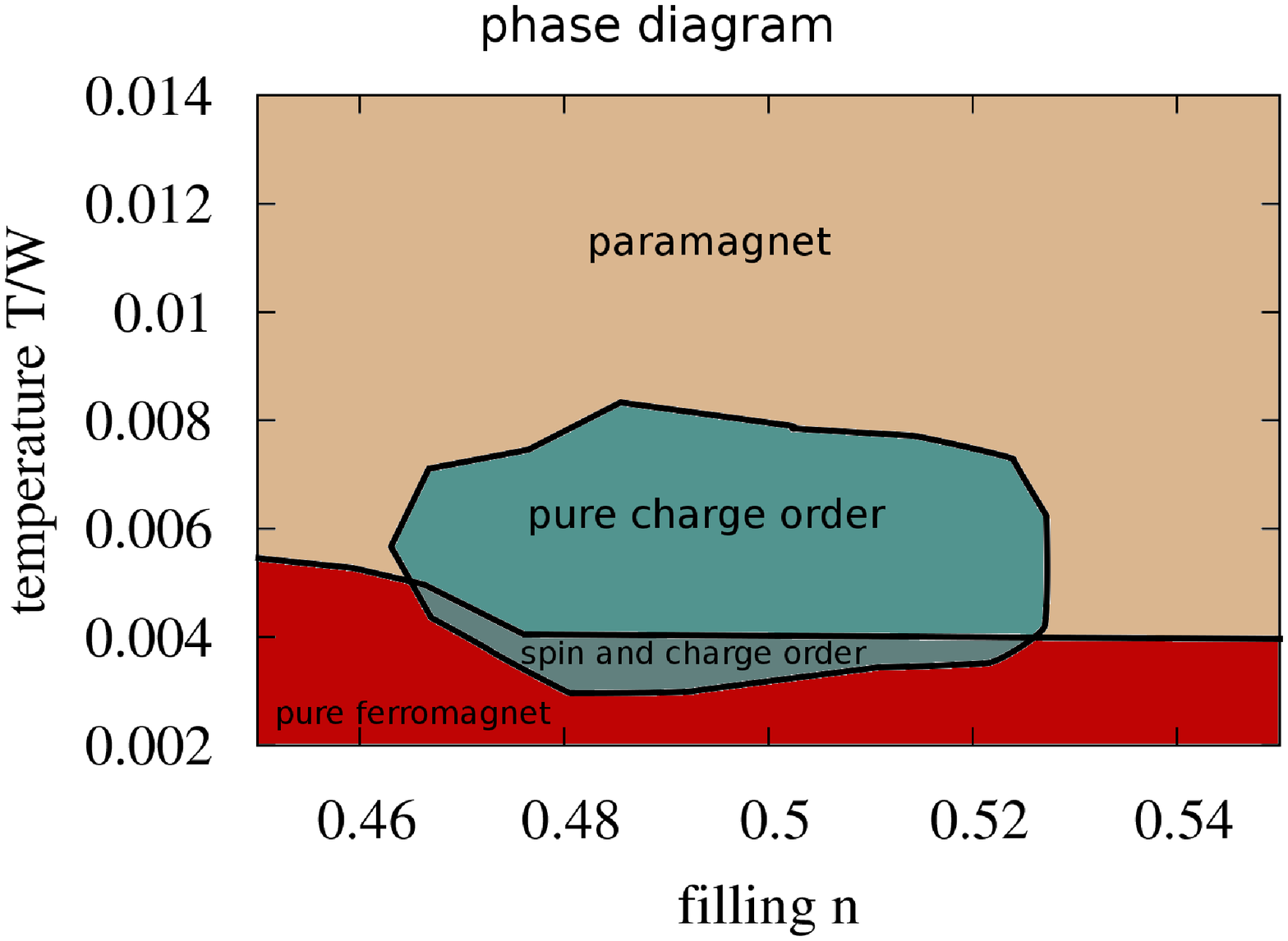}
\caption{(Color online) Spin polarization (left panel) and charge
  order-parameter (middle panel)  calculated from approximately 300
  data points for  $J/W=0.35$, simultaneously allowing for charge- and
  magnetic-order  in two-site cluster. From these data the qualitative phase
  diagram (right panel) is drawn. The phase diagram comprises a pure
  ferromagnetic state at low temperatures, a transition region where
  charge and spin order can be found, a pure charge ordered state at
  intermediate temperatures, and a paramagnetic disordered state at
  high temperatures.
 \label{fig6}}
\end{figure*}

In Fig. \ref{fig5} we show spectral properties of the CDW at
$T=0$. The spectral functions of the CDW for both 
sublattices and for all coupling strengths $J$ possess a gap at
the Fermi energy, thus being insulating. This could be expected due to
the long range order and commensurate filling. 

The qualitative shape of the spectrum can be understood in terms of a
doped Kondo insulator. At half filling the Kondo effect splits the
spectrum into two bands with a gap at the Fermi energy. For quarter
filling, the Fermi energy is located at the center of the lower
band. Because of the doubling of the unit cell due to the 
CDW, this lower band is again split into two bands with van Hove
singularities at their band edges. This qualitative behavior can be
explicitly seen for the $J/W=0.5$ curve in Fig. \ref{fig5}. The upper
band of the Kondo insulator, which is unaffected by the charge order,
is located at $\omega/W\approx 0.7$. The gap of the Kondo insulating
phase corresponds to the dip located at $\omega/W\approx 0.2$. The
lower band of the Kondo insulator, which is now split due to the CDW
with van Hove singularities at the band edges,
is located around the Fermi energy $\omega/W=0$. The charge order
strongly affects the lower band of the Kondo insulator, which is
filled differently on the two sublattices. For the smaller coupling
strength $J/W=0.2$, the gap of the Kondo insulator is strongly smeared
out, which makes the identification of these structures more difficult.

\section{competition between ferromagnetism and charge order}

As already noted in the beginning, most previous studies found
magnetism at quarter filling. Up to now we have completely suppressed
magnetic ordering. In Fig. \ref{fig6} we show the results of
calculations simultaneously allowing for charge order and
magnetism. In these calculations we allowed for any
  combination of magnetism and charge order on a two-site
  cluster. Thus, in principle also antiferromagnetic solutions would have
  been possible, but only ferromagnetic results turned out to be
  stable in our calculations.
The left panel and the middle panel show the strength of the spin
polarization and the charge order parameter, respectively. From this
we draw a qualitative phase diagram in the right panel. 

It should be
noted that for these calculations we use mixing
techniques within DMFT,\cite{zitko2009} which help improving the
convergence towards the solution.
Furthermore, there are parameter regions for which the DMFT
calculations without these mixing
techniques at $T=0$ does not seem to converge, but get stuck in
oscillating solutions. 
This could indicate that there are other competing solutions, which
however could not be stabilized.

From the panels
showing the order parameters, it is clearly 
seen that the charge order parameter vanishes again at very low
temperatures. The ground state is a  magnetic
(ferromagnetic) state without charge order. Thus, charge order and
magnetic order are  competing at $T=0$.
However, charge order
can be found when increasing the temperature. The magnetic state
vanishes at a lower temperature compared to the transition temperature
of the CDW. Therefore, exactly at quarter filling we will find a transition
from a paramagnet to a charged ordered state which is followed by a region of
coexistence between charge order and magnetism, when decreasing the
temperature. Eventually at even smaller temperatures, this coexistence
region vanishes, and a magnetic state without charge order is formed.

It should be noted that our calculations have been performed for an
infinite dimensional Bethe lattice. 
Because of the simplicity of the
Bethe lattice, i.e. there are no closed loops, order patterns 
involving more than two sites and combining magnetic- and charge-order are
unlikely. 
 Furthermore, we have also checked different
  high-dimensional lattice 
geometries like the hyper-cubic-lattice, which  exhibits a Gaussian
DOS, and we found similar results as for the Bethe
lattice. Thus, our results should be valid for bipartite lattices with similar
DOS and large coordination number. We again want to
stress that a lattice with constant DOS does not show charge
order. Therefore, a curvature within the DOS is necessary.
However, especially for
low-dimensional lattices 
like the two dimensional square lattice, more
complicated ground states allowing for a cooperation between charge
order and magnetism might be favorable. In a recent study,\cite{misawa2013} a combination of charge order and antiferromagnetism on four
site clusters have been observed. 
Similar order patterns have also
been observed for frustrated low dimensional
lattices.\cite{Motome2010,Hayami2011,Hayami2012,Ishizuka2012,Akagi2012} 
However, the existence of these states strongly depend on the lattice geometry,
and these states are not easily generalized to arbitrary lattices for
higher dimensions, $d>2$.

\section{conclusions}

We have shown that charge order can exist in the Kondo lattice model
without any repulsive interaction between different lattice sites or
between {\it c}- and {\it f}- electrons. For this purpose we have used DMFT, which
is exact in infinite dimensions, but is supposed to be a good
approximation for three dimensional systems. We have shown that a
paramagnetic charge ordered state exists for small
coupling strengths, $J<0.65$ for a Bethe 
lattice, and vanishes as a first order phase transition for
strong coupling. We have also shown that the charge ordered
state at quarter filling is an insulator, as can be expected due to
the doubling of the 
unit cell. As an explanation for the existence of charge order, 
the non-linear dependence of the Kondo  energy is proposed.

Furthermore, we have studied the competition/cooperation of charge order
and magnetic order. In these calculations we have found that the 
ferromagnetic state without charge order is the stable state at zero
temperature, and charge order can be only found for increasing 
temperature when the magnetic state vanishes. 
However, in all our
calculations we have seen that the CDW has the higher transition
temperature compared to the transition temperature of the magnetic state. Thus, the experimentally
observed charge order can be explained without any repulsive inter-site
interactions, as long as the system is in the Kondo regime (odd number
of f-electrons), as it is assumed for a Kondo lattice model.

Shortly before finishing this work, we became aware of a study by \textcite{misawa2013}, who uses cluster methods and
Variational Monte Carlo for
analyzing charge order in the Kondo lattice model. 

\begin{acknowledgments}
We acknowledge discussions with T. Yoshida and T Pruschke.
RP thanks the Japan Society for the Promotion of Science (JSPS)
for the support  through its FIRST Program. SH acknowledges the
Grant-in-Aid for JSPS Fellows. 
NK acknowledges support through KAKENHI (No. 25400366 and 22103005), the Global COE Program ``The Next Generation of Physics, Spun from
Universality and Emergence'' from MEXT of Japan, and JSPS through its
FIRST Program.  The numerical calculations were performed at the ISSP
in Tokyo and on the SR16000 at YITP in Kyoto University.
\end{acknowledgments}


\end{document}